\documentclass[pre,showpacs,preprint,superscriptaddress]{revtex4-1}
\usepackage{amssymb,graphicx,amsbsy,bm}

\begin{document}
\title{Double stochastic resonance in the mean-field $q$-state clock models}
\author{Seung Ki Baek}
\affiliation{School of Physics, Korea Institute for Advanced Study, Seoul
130-722, Korea}
\author{Beom Jun Kim}
\email[Corresponding author, E-mail: ]{beomjun@skku.edu}
\affiliation{BK21 Physics Research Division and Department of Physics,
Sungkyunkwan University, Suwon 440-746, Korea}

\begin{abstract}
A magnetic system with a phase transition at temperature $T_c$ may exhibit
double resonance peaks under a periodic external magnetic field
because the time scale matches the external frequency at two different
temperatures, one above $T_c$ and the other below $T_c$.
We study the double resonance phenomena for the mean-field $q$-state clock
model based on the heat-bath-typed master equation. We find double peaks
as observed in the kinetic Ising case ($q=2$) for all $q\ge 4$, but for the
three-state clock model ($q=3$), the existence of double peaks is possible
only above a certain external frequency since it undergoes a discontinuous
phase transition.
\end{abstract}

\maketitle

\section{Introduction}

From extensive investigations on stochastic resonance~\cite{*[{See, e.g., }] [{
for a general review on stochastic resonance.}] gamma}, 
it is now widely accepted that noise can play a constructive role.
A schematic example of a particle in a double-well potential 
illustrates that the particle can move in a synchronized way with a weak external periodic force, when its
average waiting time inside a well, determined by the noise strength,
is comparable to the half of the period of the external forcing~\cite{gamma}.
This is what is generally called a {\em time-scale matching
condition} in studies of stochastic resonance.
A particularly interesting model system for stochastic resonance is the
kinetic Ising model
since it can be regarded as coupled two-state oscillators with many
degrees of freedom under thermal
fluctuations~\cite{neda,*leung2,*brey,*neda3d,double}.
The probability to flip a spin $j$ from $S_j$ to $-S_j$ is given
by the Glauber dynamics~\cite{glauber} as
\begin{equation}
w(S_j \rightarrow -S_j) = \frac{1}{1 + \exp\{-\beta [E(S_1,
\ldots, S_j, \ldots, S_N) - E(S_1, \ldots, -S_j, \ldots,
S_N) ]\}},
\label{eq:glauber}
\end{equation}
where $\beta$ is the inverse temperature defined by $\beta \equiv 1/T$ with
the Boltzmann constant $k_B \equiv 1$, and $N$ is the system size.
The energy $E$ is a function of the spin configuration $(S_1,
\ldots, S_N)$, given by
\[
E = - J \sum_{\left< ij \right>} S_i S_j - h \sum_i S_i,
\]
where $J$ is a coupling strength, $\sum_{\left< ij \right>}$ runs over the
nearest neighbors, and $h$ is an external magnetic field.
If we ignore spin correlations and assume that each spin experiences the mean
field of the system, we arrive at the mean-field kinetic Ising model, which
has served as an ideal starting point to study exact results on 
stochastic resonance~\cite{leung,double}. This model has been analytically
shown that there can be two temperatures where the time-scale matching
condition is met, one above the critical temperature $T_c$ of the system and
the other below $T_c$. The reason for this {\em double} stochastic resonance
is that the intrinsic time scale of the system diverges in both the sides,
whether $T$ approaches $T_c$ from above or from below, so that the matching
with the external frequency can occur on either side. Recently, the similar 
mechanism of time-scale matching in quantum kinetic Ising model is shown to
be responsible for the double resonance peaks in quantum stochastic resonance~\cite{sghan}.

A natural extension of the Ising model is the $q$-state clock model,
where each spin $\theta$ has an angle among discrete values $2\pi n/q$ where
$n=0, \ldots, q-1$. The spin at the $j$th site should now take a vector form as
$\boldsymbol{S}_j = (\cos \theta_j, \sin \theta_j) = \left(\cos(2\pi n_j/q), 
\sin(2\pi n_j/q) \right)$, and the energy function is
accordingly rewritten as
\[
E = - J \sum_{\left< ij \right>} \boldsymbol{S}_i \cdot \boldsymbol{S}_j -
\boldsymbol{h} \cdot \sum_i \boldsymbol{S}_i,
\]
where $\boldsymbol{h} = (h_x, h_y)$ also takes a vector form.
The total magnetization is $\boldsymbol{M} = N^{-1} \sum_i
\boldsymbol{S}_i$, and its magnitude $M \equiv \left| \boldsymbol{M} \right|$
will be used as an order parameter.
The Ising model corresponds to the case of $q=2$, and
the $XY$ model can be studied by taking the limit of $q \rightarrow \infty$.
If we wish to construct a kinetic dynamics for this $q$-state clock model,
the probability to update spins such as Eq.~(\ref{eq:glauber}) for the Ising
case is readily obtainable by considering the heat-bath algorithm~\cite{loison}.

In this work, we check the double stochastic resonance in the $q$-state
clock model within the linear response theory. When the external field
$\boldsymbol{h}$ is parallel to the magnetization vector $\boldsymbol{M}$,
we find qualitatively the same double-resonance feature for all $q>3$.
When the field is
perpendicular to $\boldsymbol{M}$, however, the response is more
complicated, and one of the resonance peaks will disappear as we approach
the $XY$-model limit by taking $q \rightarrow \infty$. We will pay particular
attention to the case of $q=3$, because the system undergoes a discontinuous
phase transition unlike the other values of $q \ge 2$~\cite{kihara}. This
work is organized as follows. In Sec.~\ref{sec:eqmotion}, we derive the
equation of motion in terms of $M$ for the general $q$-state clock model.
From this equation of motion, we discuss the static phase transitions in
Sec.~\ref{sec:static}. Then, Sec.~\ref{sec:relax} examines responses of the
system when perturbed by a small amount from the static equilibrium. In
Sec.~\ref{sec:resonance}, we will see how the system under thermal
fluctuations responds in a resonant way when the perturbation is given as a
periodic magnetic field. Then, we conclude this work in
Sec.~\ref{sec:summary}.

\section{Equation of motion}
\label{sec:eqmotion}

The master equation describing the probability distribution function  
$P({\bm \theta};t)$ for the spin configuration ${\bm \theta}$ at time $t$
can be written as
\begin{equation}
\frac{d}{dt}P({\bm \theta}; t) = 
-\sum_{j=1}^N \sum_{\theta_j'} w_j(\theta_j \rightarrow \theta_j') P({\bm \theta};t)
+ \sum_{j=1}^N \sum_{\theta_j'} w_j(\theta_j' \rightarrow \theta_j) P({\bm \theta}';t),
\label{eq:master}
\end{equation}
where ${\bm \theta}' \equiv \{ \theta_1, \theta_2, \cdots, \theta_j', \cdots, \theta_N\}$ 
differs from ${\bm \theta} \equiv \{ \theta_1, \theta_2, \cdots, \theta_j, \cdots, \theta_N\}$ 
only at one site $j$. In the summations over $\theta_j'$, note that 
inclusion of the term for $\theta_j' = \theta_j$ gives null contribution
in total, and thus
has been included only for convenience.
In the heat-bath algorithm, the transition rate is given by 
\[
w_j (\theta_j \rightarrow \theta_j') = Z_j^{-1} \exp [\beta F_j
\cos(\theta_j' - \phi_j)],
\]
where
\[ Z_j \equiv \sum_{\theta_j} \exp [\beta F_j \cos(\theta_j - \phi_j)] \]
with the inverse temperature $\beta$. The local field is defined as
\begin{equation}
F_j e^{i\phi_j} \equiv \frac{J}{z} \sum_{k \in \Lambda_j} e^{i\theta_k} + h_j
\label{eq:F}
\end{equation}
with magnitude $F_j$ and phase $\phi_j$,
where $\Lambda_j$ is the set of nearest-neighboring sites of $j$ and $z$ is the
coordination number ($z \equiv |\Lambda_j| = {\rm const}$). We also denote the
external local field $h_j$ as a complex number so that its real (imaginary) part 
yields the field in $x$ ($y$) direction. 
The use of the heat-bath transition rate has a great benefit in calculation 
since it does not depend on the initial state, i.e., $w_j(\theta_j \rightarrow \theta_j') = w_j(\theta_j')$, 
which enables us to write the master equation~(\ref{eq:master}) as
\begin{equation}
\frac{d}{dt}P({\bm \theta}; t) = 
-\sum_{j=1}^N P({\bm \theta};t)
+ \sum_{j=1}^N \sum_{\theta_j'} w_j(\theta_j' \rightarrow \theta_j) P({\bm \theta}';t).
\label{eq:qmaster}
\end{equation}
For an arbitrary single-spin function, denoted by $f(\theta_l)$, the following
equation is derived from the master equation 
\begin{equation}
\frac{d}{dt} \left< f(\theta_l) \right> 
= -\left< f(\theta_l) \right> + 
\left<  \frac{\sum_{\theta_l}
\exp[\beta F_l \cos(\theta_l-\phi_l)]
f(\theta_l)}{\sum_{\theta_l} \exp [\beta F_l \cos(\theta_l -
\phi_l)]} \right>,
\label{eq:evolution}
\end{equation}
as explained in Appendix~\ref{app:eqm}.
When $q=2$, it recovers the kinetic Ising case in Ref.~\cite{leung}.
For the globally-coupled system with no external
field, Eq.~(\ref{eq:F}) simply corresponds to the complex magnetization
(we henceforth set $J \equiv 1$)
\begin{equation}
Me^{i\phi} = \frac{1}{N}\sum_{k=1}^N e^{i\theta_k} 
\label{eq:M}
\end{equation}
and Eq.~(\ref{eq:evolution}) turns out to be
\begin{equation}
\frac{d}{dt} \left< f(\theta_l) \right> =-\left< f(\theta_l) \right> 
+ \left<  \frac{\sum_{\theta_l}
\exp[\beta M \cos(\theta_l-\phi)]
f(\theta_l)}{\sum_{\theta_l} \exp [\beta M \cos(\theta_l -
\phi)]} \right>.
\label{eq:evolve2}
\end{equation}
We further use $f(\theta_l) = e^{i(\theta_l - \phi)}$ to get
$\left< f(\theta_l) \right> = (1/N)\left< \sum_l e^{i(\theta_l - \phi)} \right> = \langle M \rangle$ and
\begin{equation}
\frac{d\left<M\right>}{dt} = -\left<M\right> + \left< \frac{\sum_{\theta_l} e^{i(\theta_l-\phi)} \exp[\beta M
\cos(\theta_l-\phi)]}{\sum_{\theta_l} \exp[\beta M \cos(\theta_l-\phi)]} \right>.
\end{equation}
The Hamiltonian of the $q$-state clock model without external field is invariant both
under the uniform rotation, i.e., $\theta_l \rightarrow \theta_l + \phi$, and under
the reflection, i.e., $\theta_l \rightarrow -\theta_l$, which leads  to
\begin{equation}
\frac{d\left<M\right>}{dt} = -\left<M\right> + \left< \frac{\sum_{\theta_l} \cos\theta_l \exp(\beta M
\cos\theta_l)}{\sum_{\theta_l} \exp(\beta M \cos\theta_l)} \right>.
\end{equation}
For the globally-coupled system in thermodynamic limit, the mean-field
approximation becomes exact and we can drop the average symbols to get
\begin{equation}
\label{eq:mf}
\frac{dM}{dt} = -M + \frac{\sum_{\theta_l} \cos\theta_l \exp(\beta M
\cos\theta_l)}{\sum_{\theta_l} \exp(\beta M \cos\theta_l)}.
\end{equation}

\section{Equilibrium phase transition}
\label{sec:static}

\begin{table}
\caption{$A_{qm} =\sum_{n=0}^{q-1}\cos^m(2\pi n/q)$ for $m=0,1,\cdots,4$. 
We also list the critical value of the inverse temperature $\beta_c$. 
For the $q$-state globally-coupled clock model
$\beta_c = 2$ for all $q \geq 4$. See text for details.
}
\begin{tabular*}{\hsize}{@{\extracolsep{\fill}}ccccc}
\toprule
& $q=2$ & $q=3$  & $q=4$ & $q \geq 5$ \\
\colrule
$A_{q0}$ &  \multicolumn{4}{c}{$q$}  \\
$A_{q1}$ &  \multicolumn{4}{c}{$0$}  \\
$A_{q2}$ &   2 & \multicolumn{3}{c}{$q/2$}  \\
$A_{q3}$ &   0 & $3/4$ & \multicolumn{2}{c}{$0$} \\
$A_{q4}$ &   2 & $9/8$ & 4 & $3q/8$ \\
$\beta_c  $ &   1 & $8\ln 2/3 \approx 1.848~39$ & \multicolumn{2}{c}{$2$} \\
\botrule
\end{tabular*}
\label{tab:Aqm}
\end{table}

We are going to apply the above result in Sec.~\ref{sec:eqmotion} to the
$q$-state clock model, where $\theta_l = 2\pi n/q$ with $n=0, \ldots, q-1$
in Eq.~(\ref{eq:mf}).
In this section, we restrict ourselves to a static situation ($dM/dt=0$) in
the absence of the magnetic field ($h=0$). In such a static case,
Eq.~(\ref{eq:mf}) is interpreted as $ 0 = -\partial \mathcal{F} /
\partial M$ with the free energy $\mathcal{F}$, and 
assumes the following form of a self-consistent equation for $M$:
\begin{equation}
M = \frac{\sum_{n=0}^{q-1} \cos(2\pi n/q) \exp[\beta M
\cos(2\pi n/q)]}{\sum_{n=0}^{q-1} \exp[\beta M \cos(2\pi n/q)]} = 
\frac{\partial \ln Z}{\partial (\beta M) },
\label{eq:static}
\end{equation}
where
\begin{equation} 
Z \equiv  \sum_{n=0}^{q-1} \exp[\beta M \cos(2\pi n/q)] = \sum_{m=0}^\infty \frac{(\beta M)^m}{m!} A_{qm}, 
\end{equation} 
with $A_{qm} \equiv \sum_{n=0}^{q-1}\cos^m(2\pi n/q)$. 
Equation~(\ref{eq:static}) is then expanded in the power of $\beta M$ as
\begin{eqnarray*}
M & = & \frac{ A_{q1} + \beta M A_{q2} + \frac{1}{2} \beta^2 M^2 A_{q3} +  \frac{1}{6}\beta^3 M^3 A_{q4} + O(M^4)}
{A_{q0} + \beta M A_{q1} + \frac{1}{2} \beta^2 M^2 A_{q2} + \frac{1}{6}\beta^3 M^3 A_{q3} + \frac{1}{24}\beta^4 M^4 + O(M^5)} \\
&=& \frac{A_{q1}}{A_{q0}} + \frac{A_{q0}A_{q2} - A^2_{q1}}{A^2_{q0}}(\beta M)  
 + \frac{2A^3_{q1} - 3 A_{q0}A_{q1}A_{q2} + A^2_{q0}A_{q3}}{2A^3_{q0}}(\beta M)^2 \\
& & - \frac{6A^4_{q1} - 12A_{q0}A^2_{q1}A_{q2} + 4A^2_{q0}A_{q1}A_{q3} - A^2_{q0}[A_{q0}A_{q4}-3A^2_{q2}]}{6A^4_{q0}}(\beta M)^3 + O(\beta^4 M^4).
\end{eqnarray*}
In Table~\ref{tab:Aqm}, we list values of $A_{qm}$ for $m \leq 4$. Since $A_{q0} = q$ 
and $A_{q1} = 0$ for $q \geq 1$, the above expansion is further simplified to
\begin{equation}
M = \frac{A_{q2}}{q}(\beta M) + \frac{A_{q3}}{2q}(\beta M)^2 + \frac{q A_{q4} - 3A^2_{q2}}{6q^2}(\beta M)^3 + O(\beta^4 M^4). 
\end{equation}
The second term containing $M^2$ is particularly interesting, since  
it corresponds to the cubic term in $\mathcal{F}$, and thus  
is responsible for discontinuity of a phase transition~\cite{goldenfeld}. 
Its coefficient in this case, $A_{q3}$, vanishes for every integer $q > 1$ except $q=3$ 
(see Table~\ref{tab:Aqm}). 
It agrees with our expectation since
every mean-field $q$-state clock model undergoes a continuous transition
except $q=3$, which can be transformed to the mean-field three-state Potts
model with a discontinuous transition~\cite{kihara}.

\subsection{$q \neq 3$}
\label{subsec:staticneq3}
When $q \neq 3$, $A_{q3} = 0$ (see Table~\ref{tab:Aqm}) and Eq.~(\ref{eq:static}) does 
not have the cubic term:
\[
M \approx \frac{A_{q2}}{q} \beta M + \frac{q A_{q4} - 3A^2_{q2}}{6q^2}(\beta M)^3 
\]
From Table~\ref{tab:Aqm}, $A_{q2} = A_{q4} = 2$  for $q=2$, and we find
\[ M \approx \beta M - \frac{1}{3} \beta^3 M^3, \]
which yields $M \sim (\beta - \beta_c)^{1/2}$ with $\beta_c=1$.
For $q=4$, we have $A_{q2} = 2$ and $A_{q4} = 4$, which yields 
\[ M \approx \frac{1}{2}\beta M - \frac{1}{24} \beta^3 M^3, \]
resulting in $M \sim (\beta - \beta_c)^{1/2}$ with $\beta_c=2$.
This agrees with an exact relationship between $q=2$ and $4$~\cite{suzuki}.
For $q>4$, we have $A_{q2} = q/2$ and $A_{q4} = 3q/8$ and we arrive at
\[ M \approx \frac{1}{2} \beta M - \frac{1}{16} \beta^3 M^3, \]
irrespective of $q$, which means that we always find
$M \sim (\beta - \beta_c)^{1/2}$ with $\beta_c=2$ [Fig.~\ref{fig:m3}(a)].
The scaling form of $M$ with respect to the temperature also confirms
the mean-field value $1/2$ of the magnetization critical exponent, for
all values of $q$ other than three.
We finally remark that
for the $XY$ model ($q \rightarrow \infty$), we replace the summations in
Eq.~(\ref{eq:static}) by integrals as
\begin{equation}
M = \frac{\partial}{\partial (\beta M)}\ln\left(
\int d\theta e^{\beta M \cos\theta}\right) =
\frac{I_1(\beta M)}{I_0(\beta M)},
\label{eq:xystatic}
\end{equation}
where $I_n$ is the modified Bessel function~\cite{antoni}. It is known
that $\beta_c = 2$ in the limit of $q \rightarrow \infty$~\cite{antoni,smallXY}, which
is in agreement with the above conclusion of $\beta_c = 2$ for $q \geq 4$.
These are verified by numerical calculations as shown
in Fig.~\ref{fig:m3}(a), which are obtained by directly
solving Eq.~(\ref{eq:static}) in a numerical way.

\subsection{$q = 3$}
\begin{figure}
\includegraphics[width=0.45\textwidth]{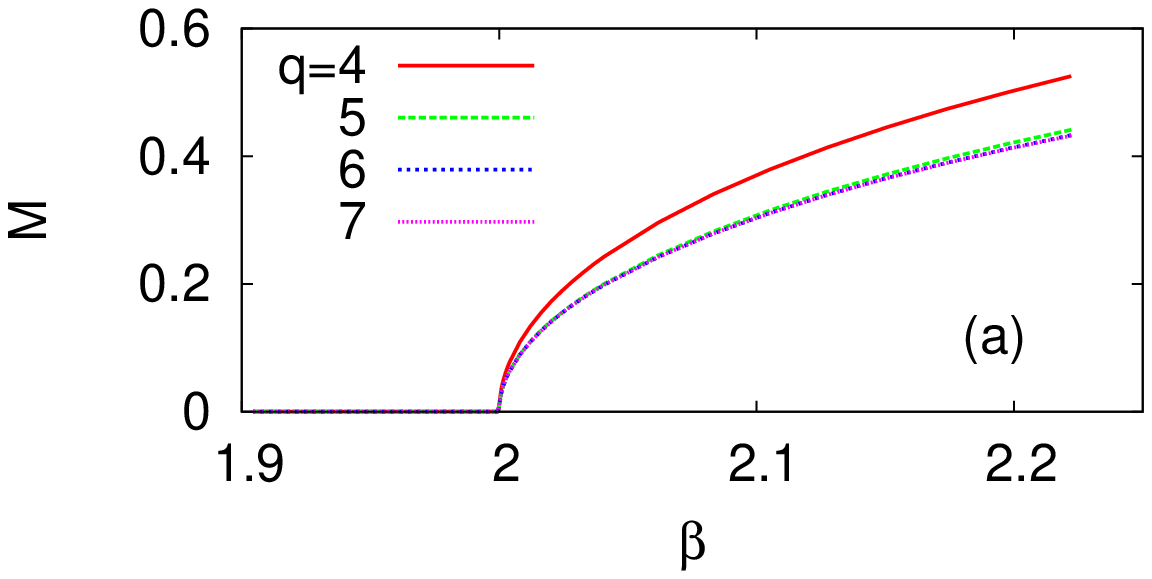}
\includegraphics[width=0.45\textwidth]{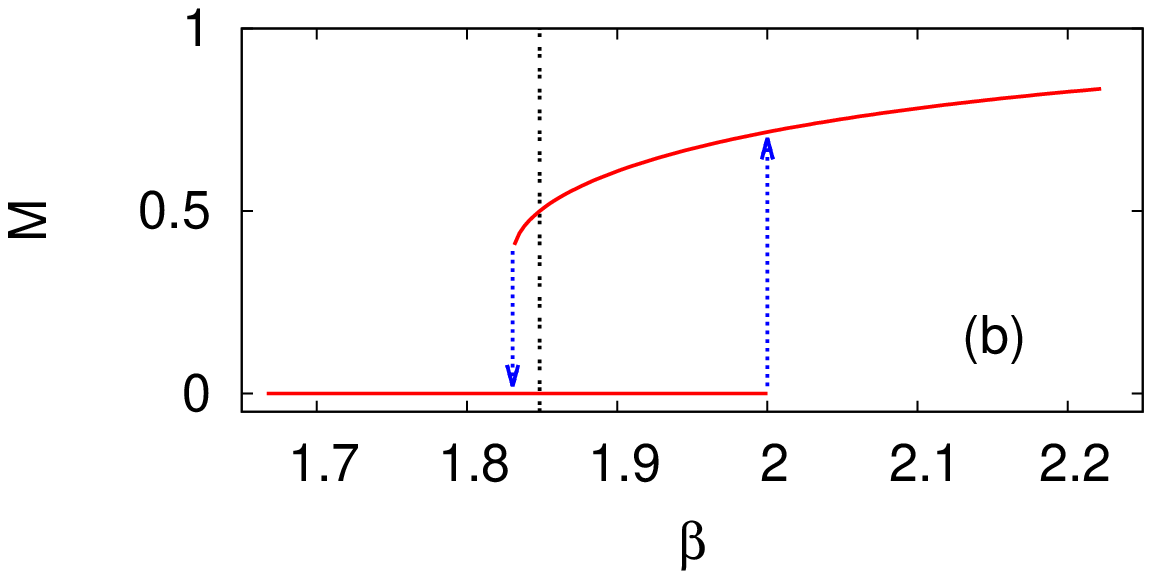}
\caption{Magnetization of the $q$-state clock model as a function of $T$.
(a) $q=4, 5, 6$, and $7$, which exhibit qualitatively the same behavior
as $M \sim |\beta - \beta_c|^{1/2}$ with $\beta_c = 2$.
(b) Three-state clock model ($q=3$), exhibiting a discontinuous transition.
The bistable region is bounded by $\beta_1 = 2$ and $\beta_2 \approx
1.830~43$ as depicted by the arrows.
The vertical dashed line indicates the transition point $\beta_c =
\frac{8}{3} \ln 2 \approx 1.848~39$.}
\label{fig:m3}
\end{figure}

\begin{figure}
\includegraphics[width=0.45\textwidth]{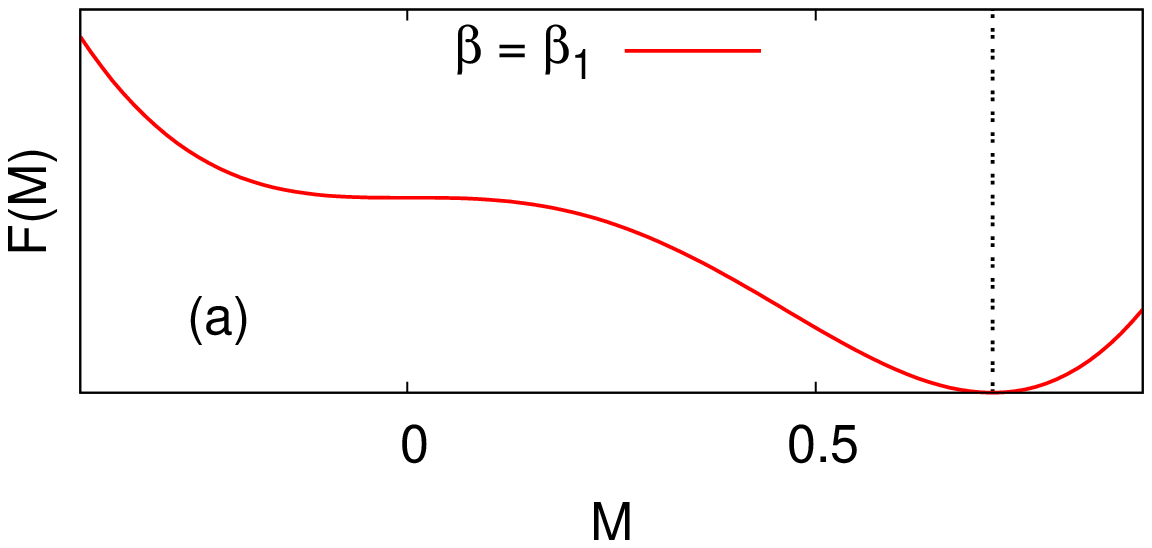}
\includegraphics[width=0.45\textwidth]{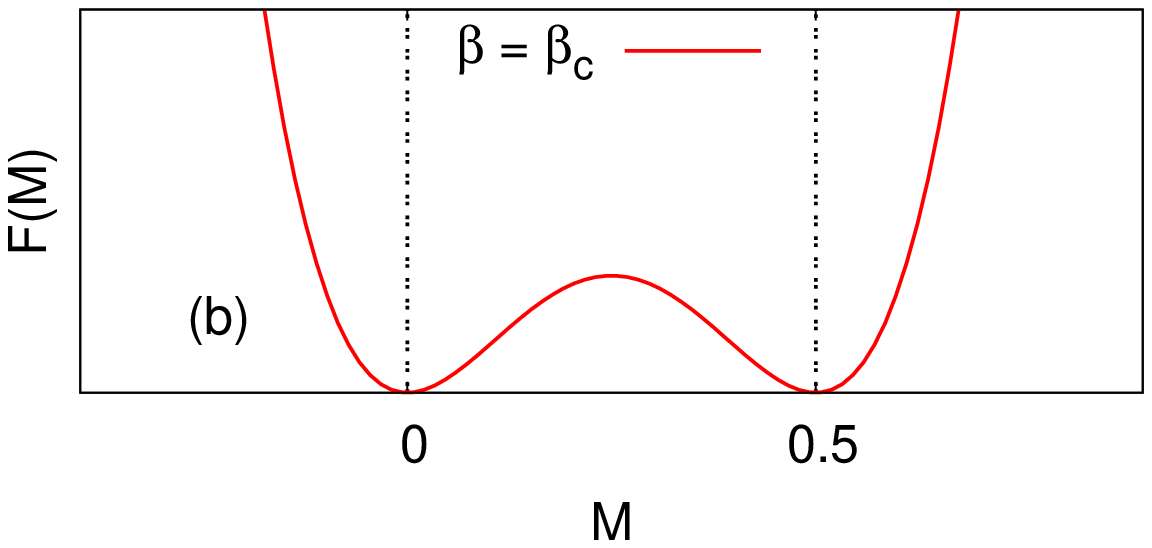}
\includegraphics[width=0.45\textwidth]{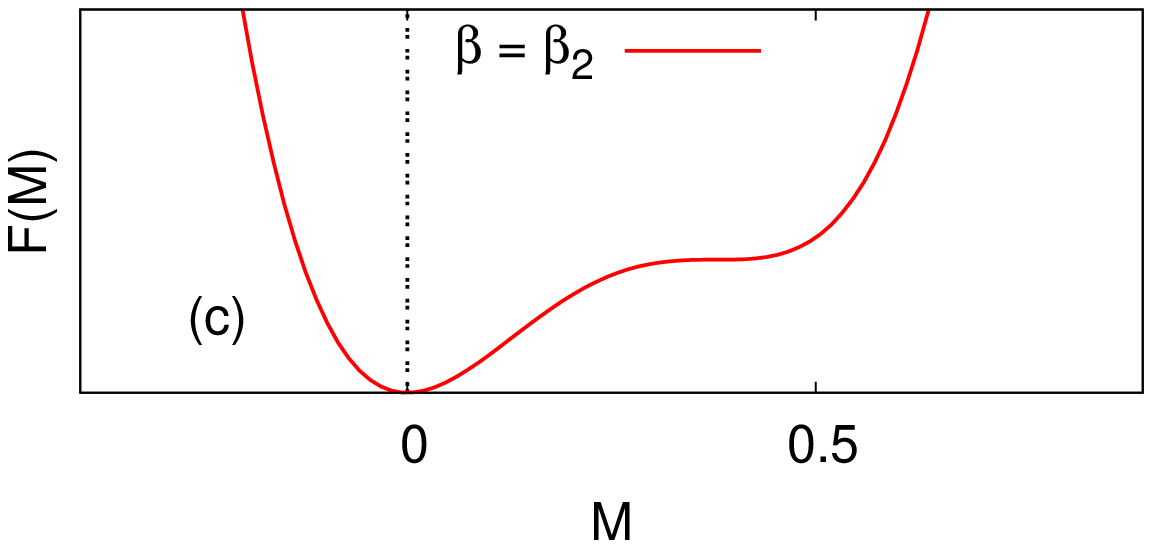}
\caption{Free-energy landscapes of the three-state clock model as a
function of $M$. The free energy  
${\mathcal F}=  M (M + 1)/2 - \beta^{-1}\log( e^{3\beta M/2} + 2 )$ is obtained from the 
integration of Eq.~(\ref{eq:q3m}).
(a) If $\beta \ge \beta_1 =  2 $, there exists only one free-energy minimum
at nonzero $M$. If $\beta$ is decreased further below $\beta_1$, 
${\mathcal F}$ begins to have two minima, but the global minimum of ${\mathcal F}$ occurs at nonzero $M$
for $\beta > \beta_c$.
(b) At $\beta = \beta_c = \frac{8}{3} \ln 2 \approx 1.848~39$, the two
free-energy minima 
have equal height, one at $M=0$ and the other at $M=1/2$. As $\beta$ is decreased
below $\beta_c$, the minimum at $M=0$ begins to locate lower than the other minimum
at nonzero $M$, but the two minima coexist until $\beta = \beta_2$ is reached.
(c) When $\beta \le \beta_2 \approx 1.830~43$, the only minimum is found
at $M = 0$. Note that the system is bistable for $\beta_2 < \beta < \beta_1$.
The vertical dashed lines indicate free-energy minima, and the
values of $M$ at these minima are observed in Fig.~\ref{fig:m3}(b).}
\label{fig:f3}
\end{figure}

In the three-state clock model, we observe a discontinuous phase transition
due to the cubic term in the expansion of the self-consistent equation~(\ref{eq:static}).
However, in order to look into the bistable region in detail, we are not allowed to use
the expansion in terms of $\beta M$, since $M$ cannot be assumed to be small in this case.
We thus start from Eq.~(\ref{eq:mf}) with the equilibrium
magnetization $M$ which should satisfy
\begin{equation}
0 = -M + \frac{e^{3\beta M/2}-1}{e^{3\beta M/2}+2} .
\label{eq:q3m}
\end{equation}
Let us assume that
there exists a certain $\beta$, where the derivative
of Eq.~(\ref{eq:q3m}) with respect to $M$ vanishes. This tells us when
bistability becomes possible.
Defining $y \equiv
e^{3\beta M/2}$, these two conditions can be written as
\begin{eqnarray*}
M &=& \frac{y-1}{y+2},\\
\beta &=& 2 y^{-1} (y+2)^2 /9,
\end{eqnarray*}
which leads to
\begin{eqnarray}
y &=& \frac{1+2M}{1-M}\nonumber,\\
\beta &=& \frac{2}{(1+2M)(1-M)}.
\label{eq:q3b}
\end{eqnarray}
Inserting these into the definition of $y$, we get a transcendental equation
for $M$ as follows:
\begin{equation}
\frac{1}{1-M} - \frac{1}{1+2M} = \ln \left(\frac{1+2M}{1-M}\right).
\label{eq:trans}
\end{equation}
Note that there is a trivial solution $M_1=0$ with $\beta_1 = 2$
from Eq.~(\ref{eq:q3b}).
One can also find a nontrivial solution of Eq.~(\ref{eq:trans}) numerically
as $M_2 \approx 0.377~201$, and the
corresponding inverse temperature is $\beta_2 \approx 1.830~43$.
It means that
the system is bistable between $\beta_1$ and $\beta_2$
[Fig.~\ref{fig:m3}(b)].
The transition point $\beta_c$ can be determined by checking when the two
free-energy minima have an equal height as in the Maxwell construction.
Interpreting the left-hand side of
Eq.~(\ref{eq:q3m}) as $\partial \mathcal{F} / \partial M$, we find that $\beta_c
 = \frac{8}{3}\ln 2 \approx 1.848~39$ (Fig.~\ref{fig:f3}), which is between
$\beta_1$
and $\beta_2$ and in agreement with Refs.~\cite{kihara,mittag}. One can also
readily check that the nonzero magnetization at $\beta_c$ is $M = 1/2$ by
using Eq.~(\ref{eq:q3m}).

\section{Relaxation time}
\label{sec:relax}

Returning back to the general $q$-state case, the evolution equation of the
magnetization vector ($M_x$, $M_y$)
is given by putting $f(\theta_l) = \cos\theta_l$ and
$f(\theta_l) = \sin\theta_l$ in Eq.~(\ref{eq:evolve2})
since $M_x = \left< \cos\theta_l \right>$
and $M_y = \left< \sin\theta_l \right>$. Within the mean-field 
scheme with no external field,  we find
\begin{eqnarray*}
\frac{\displaystyle dM_x}{\displaystyle dt} = -M_x + \frac
{\sum_\theta \cos\theta \exp[\beta M \cos(\theta - \phi)]}
{\sum_\theta \exp[\beta M \cos(\theta - \phi)]}\\
\frac{\displaystyle dM_y}{\displaystyle dt} = -M_y + \frac
{\sum_\theta \sin\theta \exp[\beta M \cos(\theta - \phi)]}
{\sum_\theta \exp[\beta M \cos(\theta - \phi)]},
\end{eqnarray*}
where $(M_x, M_y) = (M \cos \phi, M \sin \phi)$ from Eq.~(\ref{eq:M})
and $\sum_\theta$ runs over
$\theta = 2\pi n/q$ with $n=1,\ldots,q-1$.
We then add perturbation $\boldsymbol{\delta} = (\delta_x, \delta_y)$
around the equilibrium magnetization with the assumption $\beta
|\boldsymbol{\delta}| \ll 1$.
By expanding the exponential functions and leaving linear terms with
respect to $\beta \delta_x$ or $\beta \delta_y$, one can
study linear responses of the system.
Without loss of generality, we may set the initial magnetization
along the $x$-axis, i.e., $\boldsymbol{M}^\ast = (M_x^\ast, M_y^\ast) =
(M^\ast, 0)$, where $M^\ast$ satisfies Eq.~(\ref{eq:static}) as
\begin{eqnarray*}
M^\ast &=& \frac{ \sum_\theta \cos\theta \exp(\beta M^\ast \cos\theta)}
{\sum_\theta \exp(\beta M^\ast \cos\theta)},\\
0 &=& \frac{ \sum_\theta \sin\theta \exp(\beta M^\ast \cos\theta)}
{\sum_\theta \exp(\beta M^\ast \cos\theta)}.
\end{eqnarray*}
The linearized equations for perturbation $\delta_x(t) = M_x(t) -
M_x^{\ast}$ and $\delta_y(t) = M_y(t) - M_y^{\ast}$ are derived in
Appendix~\ref{app:delta} as follows:
\begin{eqnarray}
\frac{d\delta_x}{dt} &=& -\left( 1 + \beta {M^\ast}^2 - \beta C \right) 
\delta_x,\label{eq:deltax}\\
\frac{d\delta_y}{dt} &=& -\left( 1 - \beta + \beta C \right)\delta_y,
\label{eq:deltay}
\end{eqnarray}
where
\begin{equation}
C \equiv \frac{\sum_\theta \cos^2\theta \exp(\beta M^\ast \cos\theta)}
{\sum_\theta \exp(\beta M^\ast \cos\theta)}.
\label{eq:c}
\end{equation}
In other words, we have two relaxation times $\tau_{\parallel}$ and $\tau_{\perp}$ in the
parallel and perpendicular directions with respect to the equilibrium
magnetization vector, respectively, which are given as
\begin{eqnarray}
\tau_{\parallel}^{-1} &=& 1 + \beta {M^\ast}^2 - \beta C \label{eq:itaux} , \\
\tau_{\perp}^{-1} &=& 1 - \beta + \beta C = 2 + \beta({M^\ast}^2 - 1) -
\tau_{\parallel}^{-1} . \label{eq:itauy}
\end{eqnarray}
Very near to the critical point, we may assume that $M^\ast \approx 0$,
and we get $C = \sum_\theta \cos^2 \theta/\sum_\theta 1 = A_{q2}/A_{q0}$
(see Table~\ref{tab:Aqm}). When $q=2$, the $y$ component of magnetization
is not defined, and we have $C=A_{22}/A_{20} = 1$ from Table~\ref{tab:Aqm},
yielding $\tau_{\parallel}^{-1} = 1 - \beta$. For $q > 2$, we instead have $C = 1/2$ from Table~\ref{tab:Aqm}, and 
$\tau_{\parallel}^{-1} = \tau_{\perp}^{-1} = 1 - \beta/2$. Consequently, the divergence
of the relaxation time occurs when $\beta = 1$ for $q=2$, and when $\beta = 2$
for $q > 2$ ($q \neq 3$), in complete agreement with $\beta_c = 1$ for $q=2$ and
$\beta_c = 2$ for $q > 2$ (except for $q=3$ in which $M^\ast \approx 0$
is not justified) found in Sec.~\ref{subsec:staticneq3}. Also, the divergence of the relaxation time
in the form of $\tau \sim |\beta - \beta_c|^{-1}$, determines the
dynamic critical exponent $z = 2$ in the scaling form $\tau \sim \xi^z$,
since the correlation-length ($\xi$)  exponent $\nu = 1/2$ in $\xi \sim |\beta - \beta_c|^{-\nu}$
for the mean-field universality class.

\subsection{$q \neq 3$}

\begin{figure}
\includegraphics[width=0.45\textwidth]{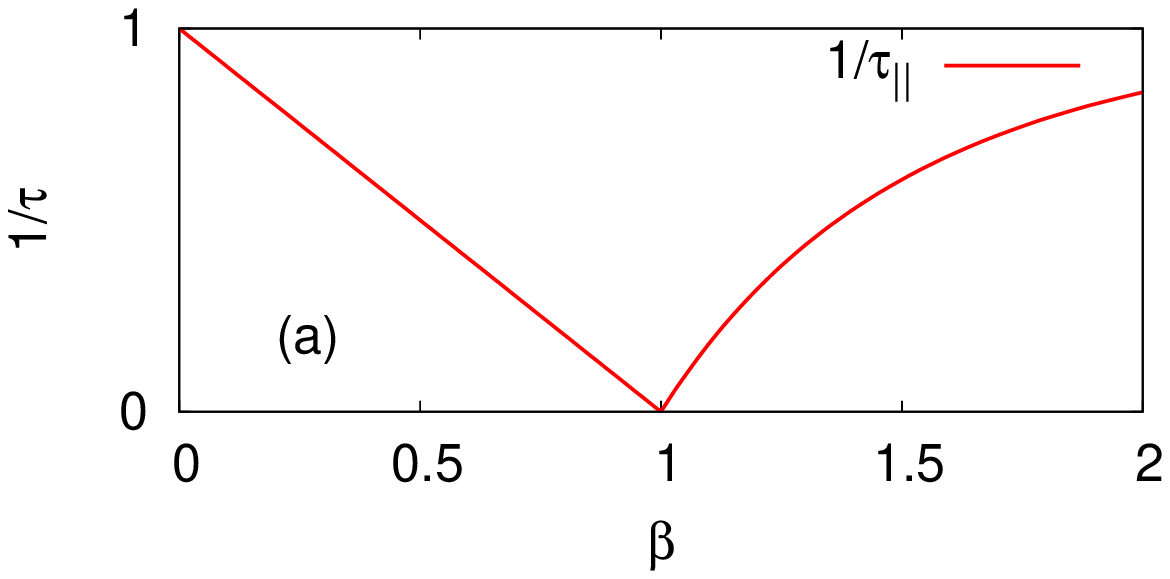}
\includegraphics[width=0.45\textwidth]{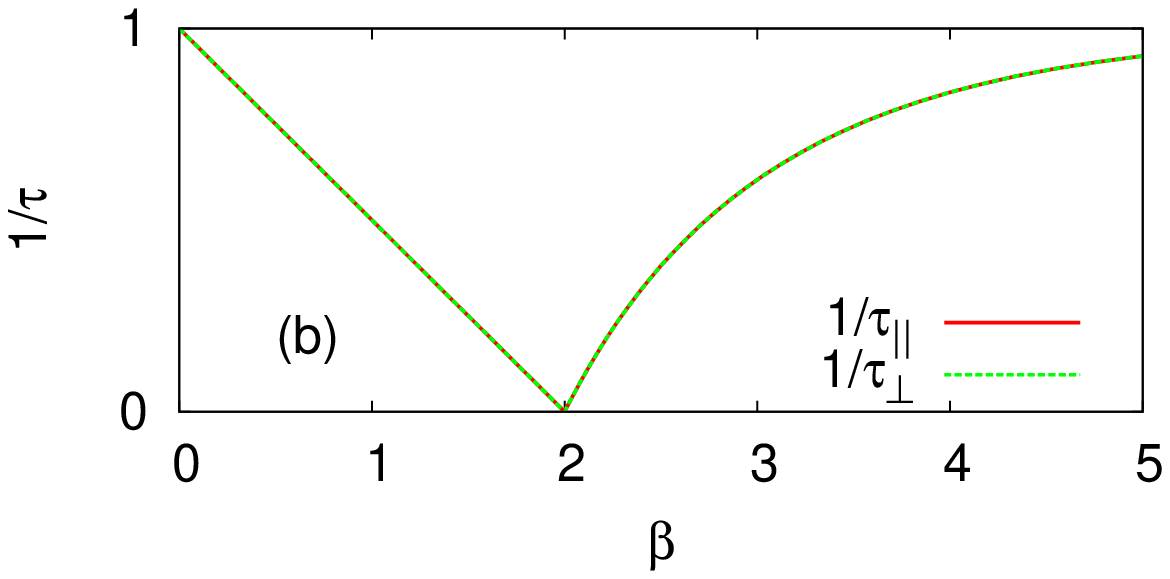}
\includegraphics[width=0.45\textwidth]{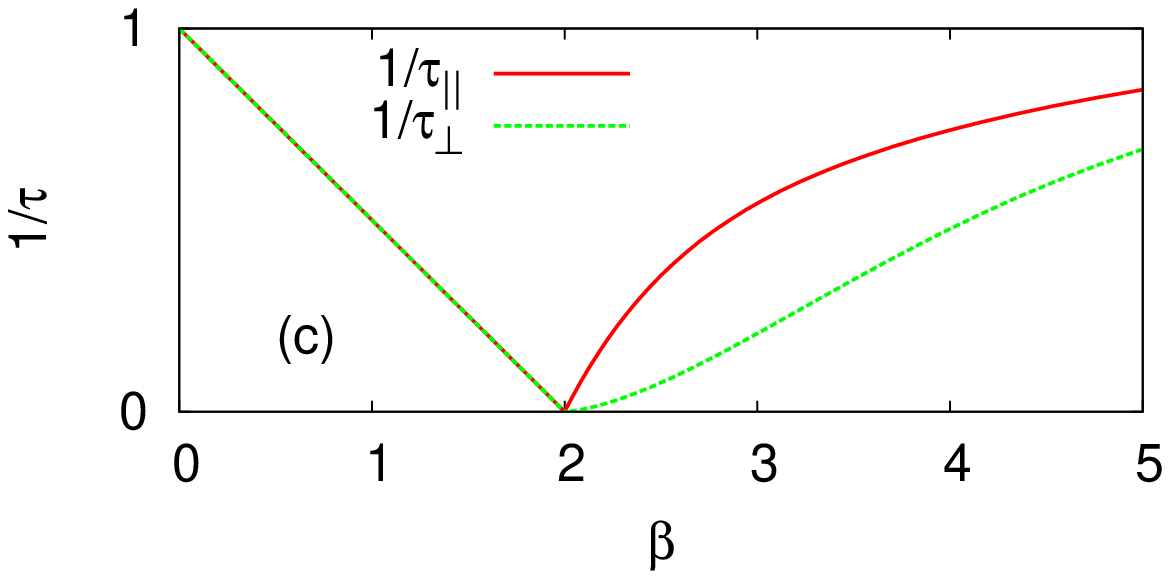}
\includegraphics[width=0.45\textwidth]{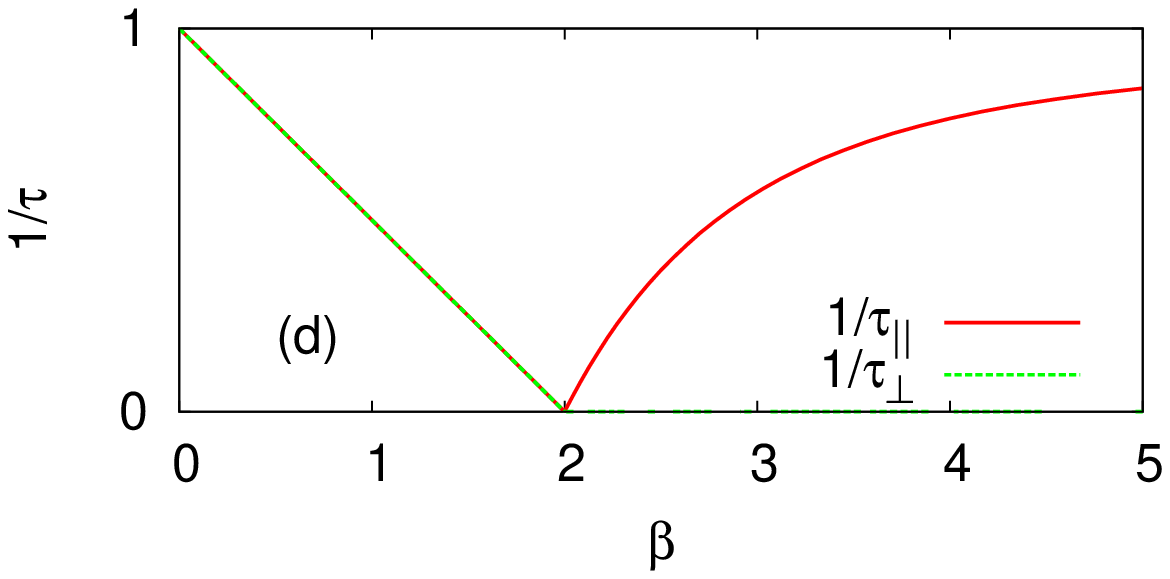}
\caption{Inverse relaxation times for the $q$-state clock models. (a) $q=2$,
(b) $q=4$, (c) $q=5$, and (d) the $XY$ limit ($q \rightarrow \infty$) where
$\tau_{\perp}^{-1} \rightarrow 0$ above $\beta_c = 2$.}
\label{fig:itau}
\end{figure}

As mentioned above,
$\tau_{\perp}$ becomes meaningless for the Ising case ($q=2$), and we get $\tau_{\parallel}^{-1} =  
1 + \beta {M^\ast}^2 - \beta$, recovering the result in
Refs.~\cite{leung,double} [Fig.~\ref{fig:itau}(a)].
For different values of $q$, we plot Eqs.~(\ref{eq:itaux}) and (\ref{eq:itauy})
in Fig.~\ref{fig:itau} by
solving Eq.~(\ref{eq:static}). For $q=4$, we find $\tau_{\parallel} =
\tau_{\perp}$ in the entire region of $T$ [Fig.~\ref{fig:itau}(b)] since
\[
C = \frac{e^{\beta M^\ast} + e^{-\beta M^\ast}}{e^{\beta M^\ast} +
e^{-\beta M^\ast} + 2}
= 1 + \left( \frac{e^{\beta M^\ast}-1}{e^{\beta M^\ast}+1} \right)^2 = 1 +
{M^\ast}^2.
\]
For $q>4$, we instead observe $\tau_{\perp} \ge \tau_{\parallel}$ at $\beta
> \beta_c$
[Fig.~\ref{fig:itau}(c)] and the difference becomes pronounced as $q$
increases. In the limiting case of $q \rightarrow
\infty$, $\tau_{\perp}$ diverges at $\beta > \beta_c$, which reflects the
$U(1)$ symmetry of
the system. This shows the following identity of the modified Bessel
function for $\beta \ge 2$,
\[
C
= \frac{1}{2} \left[ 1 +
\frac{I_2(\beta M^\ast)}{I_0(\beta M^\ast)} \right] = 1 - \frac{1}{\beta}
\]
where $M^\ast$ satisfies Eq.~(\ref{eq:xystatic}).
This observation is also related to the idea in Ref.~\cite{quasi}
that fluctuations in the angular direction can distinguish the
discrete symmetry in the clock model from the continuous symmetry of the $XY$
model.

\subsection{$q=3$}
\begin{figure}
\includegraphics[width=0.45\textwidth]{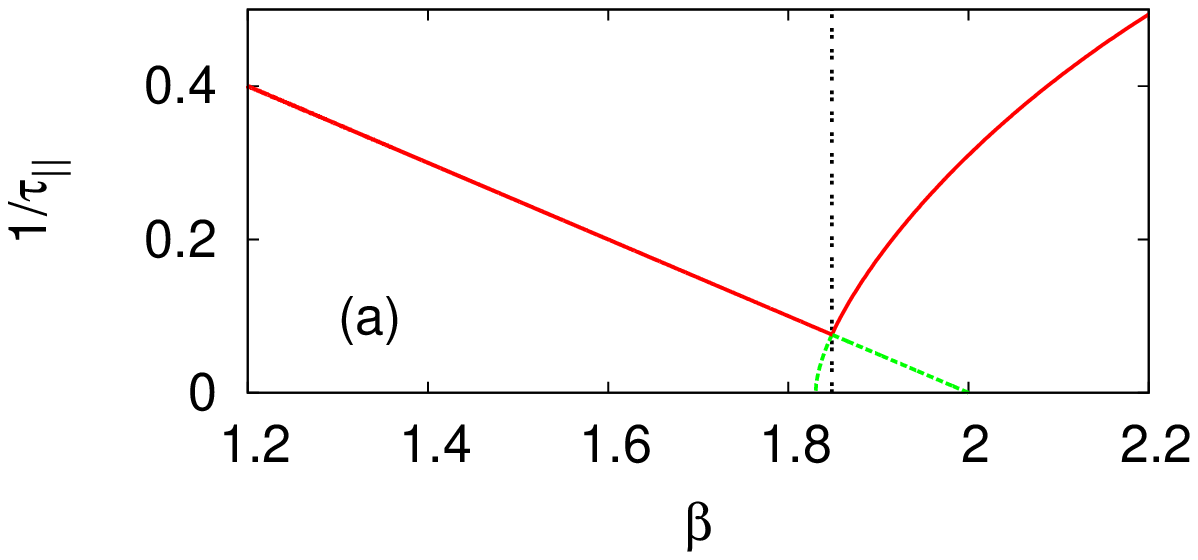}
\includegraphics[width=0.45\textwidth]{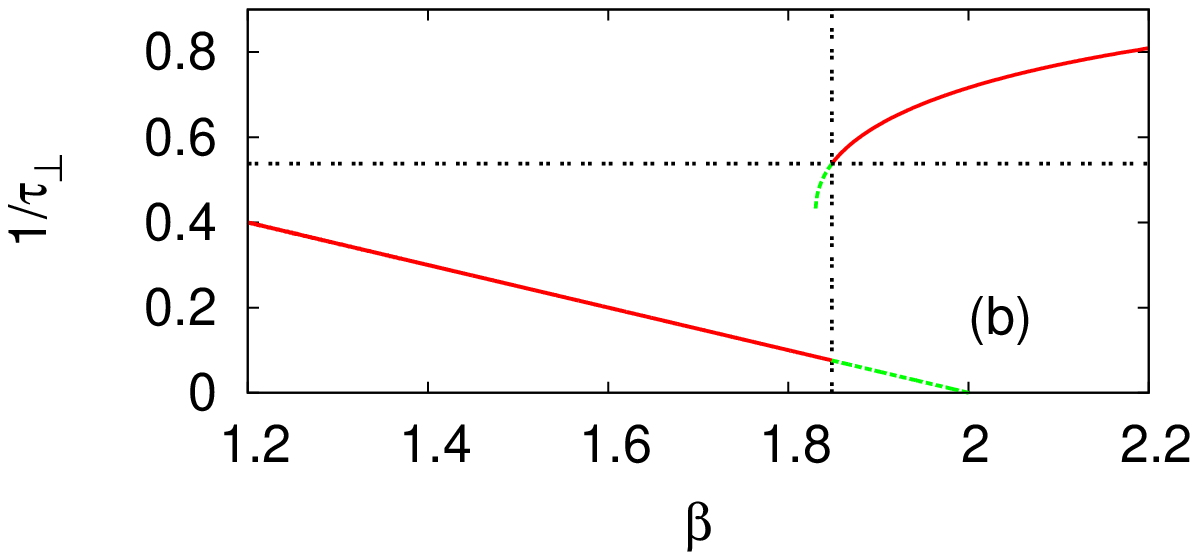}
\caption{Three-state clock model. 
(a) Inverse relaxation time $\tau_{\parallel}^{-1}$ in the direction of
magnetization and (b) $\tau_{\perp}^{-1}$ in the perpendicular direction.
The solid lines represent values at the lowest free energy,
while the dotted lines mean what one can observe at metastable states. The
vertical dashed lines indicate the transition point $\beta_c =
\frac{8}{3} \ln 2 \approx 1.848~39$. The horizontal dashed line shows
$1-\frac{2}{3}\ln 2 \approx 0.537~902$.}
\label{fig:itauy3}
\end{figure}

As explained above, this system has a bistable region between $\beta_1 = 2$
and $\beta_2 \approx 1.830~43$. Even though
the relaxation times can be expressed  in the same way,
one should note that they are defined in the linear-response
regime. In other words, even if the system becomes bistable, the relaxation
here means returning back to the original state and not jumping to the other
state.
If $M^\ast=0$, we have seen that $\tau_{\parallel}^{-1} = \tau_{\perp}^{-1} =
1-\beta/2$. So both of them
vanish if $\beta \rightarrow \beta_1^-$ from below with keeping $M^\ast=0$.
If $\beta$ approaches $\beta_2$ from above with keeping $M^\ast \neq 0$,
on the other hand, we find
\[ \tau_{\parallel}^{-1}(\beta \rightarrow \beta_2^+) = 1 - \frac{\beta_2}{2} -
\frac{\beta_2 M_2}{2} + \beta_2 M_2^2 = 0, \]
by using Eq.~(\ref{eq:q3b}). Here $M_2$ means the nontrivial solution
of Eq.~(\ref{eq:trans}). These results are plotted in
Fig.~\ref{fig:itauy3}(a). One can clearly see why those metastable branches
cannot be sustained beyond $\beta_1$ and $\beta_2$, respectively. The other
inverse relaxation time $\tau_{\perp}^{-1}$ is plotted in
Fig.~\ref{fig:itauy3}(b).
Note the jump at $\beta_c$ from $\tau_{\perp}^{-1} (\beta \rightarrow \beta_c^-) = 1
- \beta_c/2 = 1 - \frac{4}{3}\ln 2 \approx 0.075~803~8$ to $\tau_{\perp}^{-1} (\beta
\rightarrow \beta_c^+) = 1 - \frac{2}{3}\ln 2 \approx 0.537~902$. The latter
value is obtained by using Eq.~(\ref{eq:itauy}) with $M^{\ast} = 1/2$. This
behavior is in accordance with the general tendency that $\tau_{\perp}$
increases in the ordered phases as $q$ becomes larger, but manifests itself
in a discontinuous way.

\section{Resonance}
\label{sec:resonance}

If there exists a uniform external
field $\boldsymbol{h} = (h_x, h_y)$, the evolution equations are generalized
to
\begin{eqnarray}
\frac{\displaystyle dM_x}{\displaystyle dt} = -M_x + \frac
{\sum_\theta \cos\theta \exp[\beta F \cos(\theta - \phi)]}
{\sum_\theta \exp[\beta F \cos(\theta - \phi)]}\label{eq:dx}\\
\frac{\displaystyle dM_y}{\displaystyle dt} = -M_y + \frac
{\sum_\theta \sin\theta \exp[\beta F \cos(\theta - \phi)]}
{\sum_\theta \exp[\beta F \cos(\theta - \phi)]},\label{eq:dy}
\end{eqnarray}
where $(F \cos \phi, F \sin \phi) = (M_x + h_x, M_y + h_y)$ 
[see Eq.~(\ref{eq:F}) for globally-coupled ($z = N$) model with a uniform external
field]. The magnetization is again decomposed into 
$(M_x, M_y) = (M_x^\ast + \delta_x, M_y^\ast + \delta_y)$, where we set again
$M_x^\ast = M^\ast$ and $M_y^\ast = 0$ as before.
Let us furthermore assume that $\beta |\boldsymbol{h}| \ll 1$ and expand the
equations up
to the linear order. The derivation is almost the same as what we did for
Eqs.~(\ref{eq:deltax}) and (\ref{eq:deltay}), except that we have to replace
$\delta_x$ by $\delta_x + h_x$ as well as
$\delta_y$ by $\delta_y + h_y$
in expanding the second terms of Eqs.~(\ref{eq:dx}) and (\ref{eq:dy}).
This results in
\begin{eqnarray*}
\frac{d\delta_x}{dt} &=& - \delta_x - \beta\left({M^\ast}^2 - C \right) (\delta_x + h_x)\\
&=& -\tau_{\parallel}^{-1}(\delta_x + h_x) + h_x\\
&=& -\tau_{\parallel}^{-1}\delta_x + (1-\tau_{\parallel}^{-1})h_x,\\
\frac{d\delta_y}{dt} &=& - \delta_y + \beta\left(1 - C \right)(\delta_y +
h_y)\\
&=& -\tau_{\perp}^{-1}(\delta_y + h_y) + h_y\\
&=& -\tau_{\perp}^{-1}\delta_y + (1-\tau_{\perp}^{-1})h_y,
\end{eqnarray*}
where $\tau_{\parallel}^{-1}$ and $\tau_{\perp}^{-1}$ are given by Eqs.~(\ref{eq:itaux}) and
(\ref{eq:itauy}).
Since these equations have basically the same form,
we may drop the subscript $\parallel$ or $\perp$ in finding the formal solution.
If we set $h = h_0 \cos\omega t$, the resulting equation is the following:
\begin{equation}
\frac{d\delta}{dt} = -\tau^{-1} \delta + (1-\tau^{-1})h_0 \cos\omega t.
\label{eq:linear}
\end{equation}
Its solution is obtained by assuming
\[\delta(t) = \delta_0 \cos(\omega t - \sigma),\]
which results in phase shift $\sigma = \arctan(\omega \tau)$ and amplitude
\[ \delta_0 = \frac{(\tau-1) h_0}{\sqrt{1+\omega^2 \tau^2}}. \]
We can interpret this solution as follows.
When $\tau \sim O(1)$, the left-hand side of Eq.~(\ref{eq:linear}) is
negligible compared to the first term of the right-hand side since $\omega
\ll 1$ by assumption. Then, $\delta$ becomes directly proportional to $h_0
\cos\omega t$ without any phase shift, even though the amplitude will be small.
On the other hand, if $\tau \gg 1$, the first term on the right-hand side of
Eq.~(\ref{eq:linear}) becomes negligible so that $d\delta / dt$ is
proportional to $h_0 \cos\omega t$, which leads to a phase shift of
$\pi/2$ in $\delta(t)$ compared to the magnetic field.

The ac susceptibility components are defined as
\begin{eqnarray*}
\chi' &=& \frac{1}{\pi h_0} \int_0^{2\pi} d(\omega t) ~M \cos\omega t,\\
\chi'' &=& \frac{1}{\pi h_0} \int_0^{2\pi} d(\omega t) ~M \sin\omega t.
\end{eqnarray*}
The former one $\chi'$ is closely related to the occupancy ratio used in
Ref.~\cite{double} to measure how many spins are aligned in the direction of
the external field.
By inserting $\delta(t)$ here, we obtain
\begin{eqnarray}
\chi' &=& \frac{\tau-1}{1+\omega^2 \tau^2}\label{eq:chi1},\\
\chi'' &=& \frac{\omega \tau(\tau-1)}{1+\omega^2 \tau^2} = \omega \tau \chi',
\label{eq:chi2}
\end{eqnarray}
where the static value $M^{\ast}$ integrates out to zero when multiplied by
the sinusoidal functions.
When $M^\ast = 0$ and $\tau^{-1} = 1 - \beta/2$, the maximum of
Eq.~(\ref{eq:chi1}) is found at $\beta = 2\omega^2 + 2 - 2\omega
\sqrt{\omega^2+1}$, which approaches unity as $\omega \rightarrow \infty$.
In general, the extremum condition of $\chi'$ with respect to $\beta$ is
equivalent to $2 \omega \chi'' = 1$ since
\[
\frac{d\chi'}{d\beta} = \frac{d\tau/d\beta}{1+\omega^2\tau^2} -
\frac{\tau-1}{(1+\omega^2\tau^2)^2} 2\omega^2\tau\frac{d\tau}{d\beta}
= \left(\frac{d\tau}{d\beta}\right) \frac{1 -
2\omega\chi''}{1+\omega^2\tau^2}.
\]
This condition yields a solution $\omega = (\tau^2 - 2\tau)^{-1/2}$, which
can be expanded as $\tau^{-1} + \tau^{-2} + \frac{3}{2}\tau^{-3} + \cdots$.
Therefore, the optimal frequency $\omega$ for resonance coincides with
$\tau^{-1}$ to the leading order when $\tau \gg 1$. Consequently, 
the stochastic resonance occurs when the extrinsic time scale $1/\omega$
matches with the intrinsic one $\tau$~\cite{gamma,double}. 

\subsection{$q \neq 3$}

\begin{figure}
\includegraphics[width=0.45\textwidth]{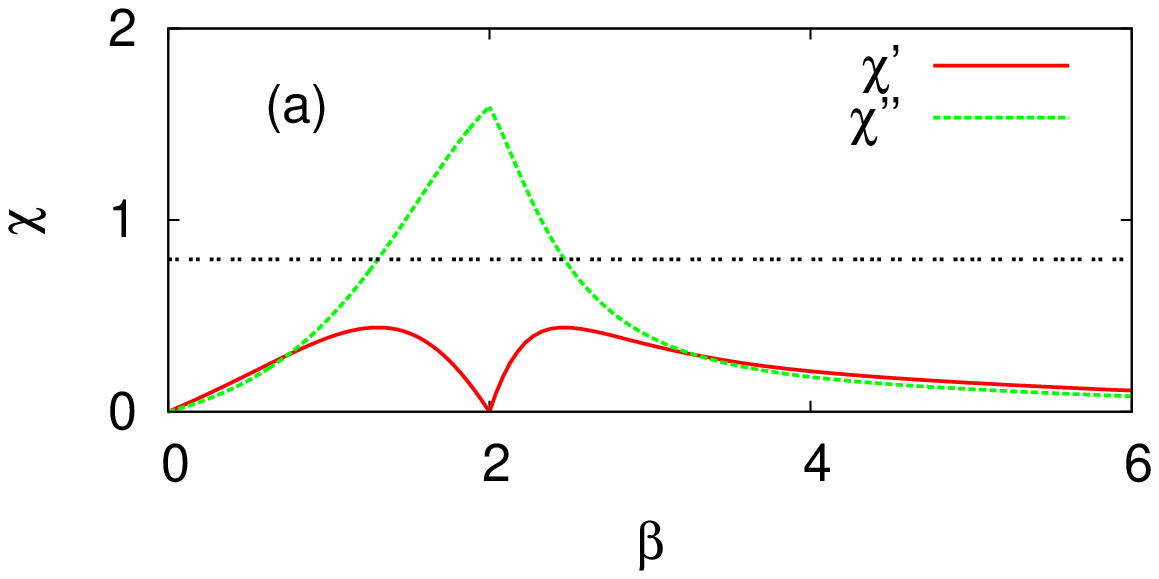}
\includegraphics[width=0.45\textwidth]{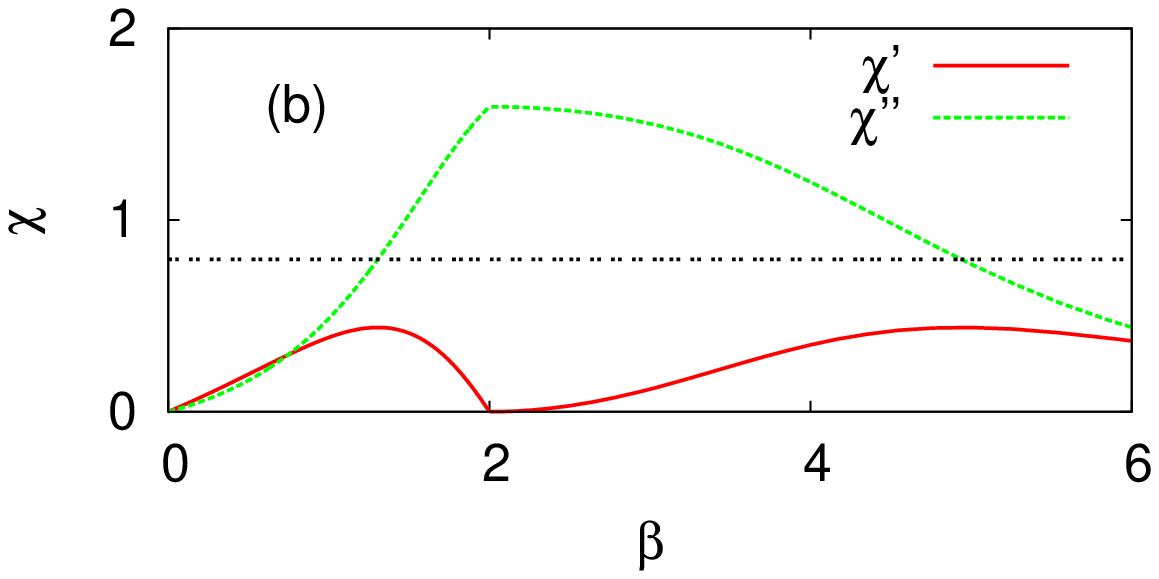}
\includegraphics[width=0.45\textwidth]{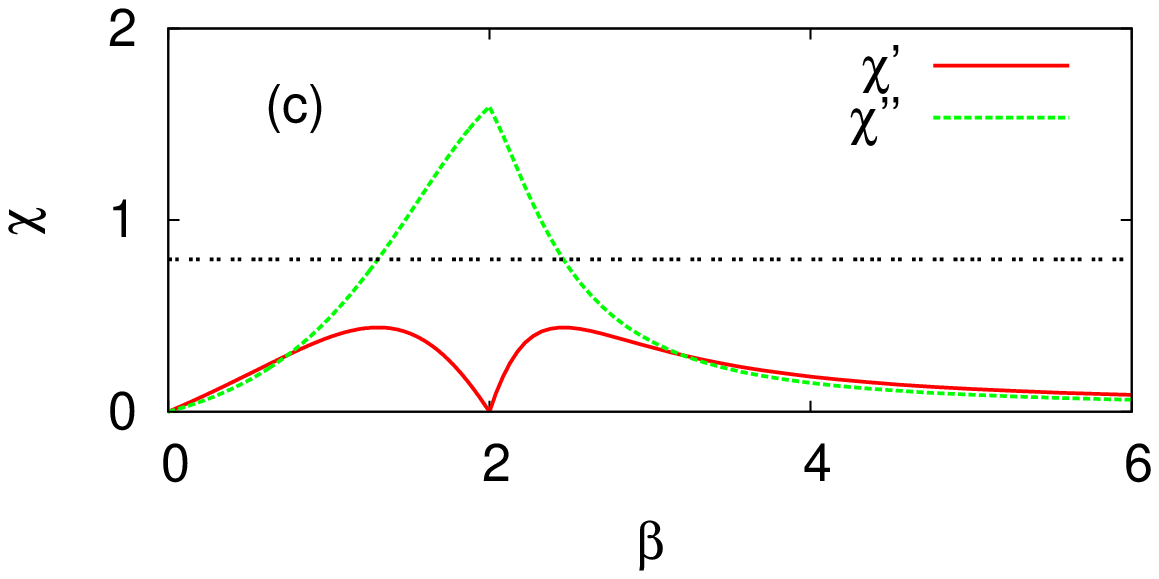}
\includegraphics[width=0.45\textwidth]{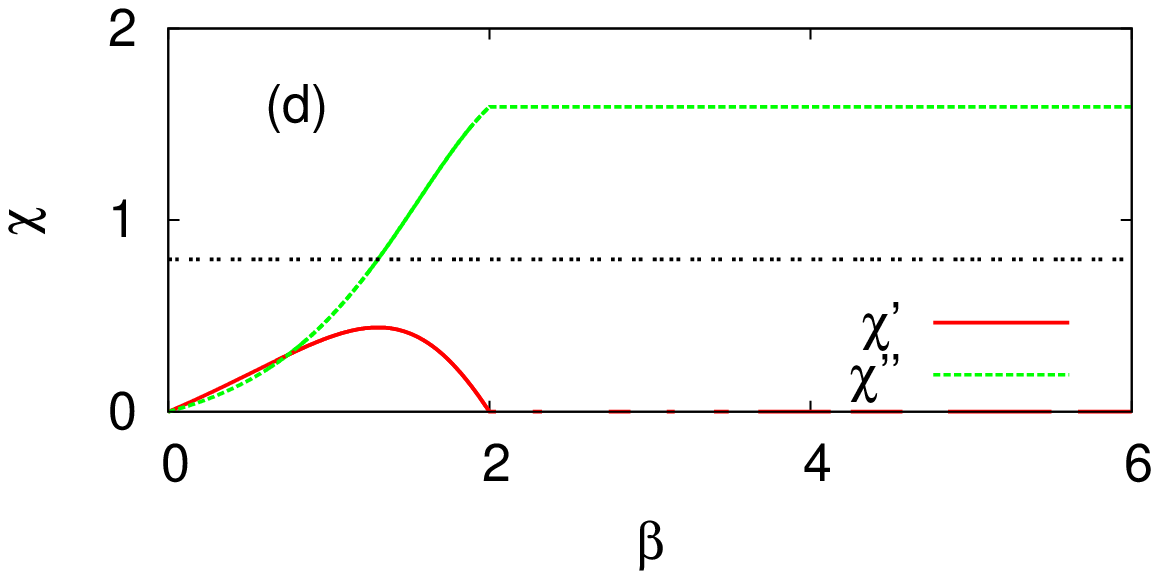}
\caption{
ac susceptibility components for $q=6$ given by
Eqs.~(\ref{eq:chi1}) and (\ref{eq:chi2}) with $\omega = 2\pi \times
10^{-1}$. (a) The field can be exerted in the $x$ direction or (b) in the
$y$ direction. The next two panels show the same plots for $q \rightarrow
\infty$, with (c) the field in the $x$ direction and (d) in the $y$ direction.
The horizontal dashed lines indicate $\chi'' = (2\omega)^{-1}$ to locate the
maxima in $\chi'$.
}
\label{fig:chi}
\end{figure}

First, we consider the field in the $x$ direction, parallel to the
magnetization. We thus use $\tau_{\parallel}$ in
place of $\tau$ in Eqs.~(\ref{eq:chi1}) and (\ref{eq:chi2}).
If the system undergoes a continuous phase transition with $q \neq 3$, one
can clearly see two peaks in $\chi'$ above and below $\beta_c$. We show the case
of $q=6$ in Fig.~\ref{fig:chi}(a), noting that
qualitatively the same behavior is observed for other $q$ values.
Now let us apply the field in the $y$ direction for $q \neq 2$ (note that
the field in $y$ direction for $q=2$ is meaningless).
When $\beta < \beta_c$, the disordered phase of the system is isotropic so
we observe the same response as above, although the field direction
has changed.
However, when $\beta > \beta_c$, the
peak is suppressed to a higher $\beta$ [Fig.~\ref{fig:chi}(b)].
As $q \rightarrow \infty$, $\chi'$ eventually vanishes and $\chi''$
becomes constant at $\beta>\beta_c$, which signals the $U(1)$ symmetry
[Fig.~\ref{fig:chi}(d)].

\subsection{$q=3$}

\begin{figure}
\includegraphics[width=0.45\textwidth]{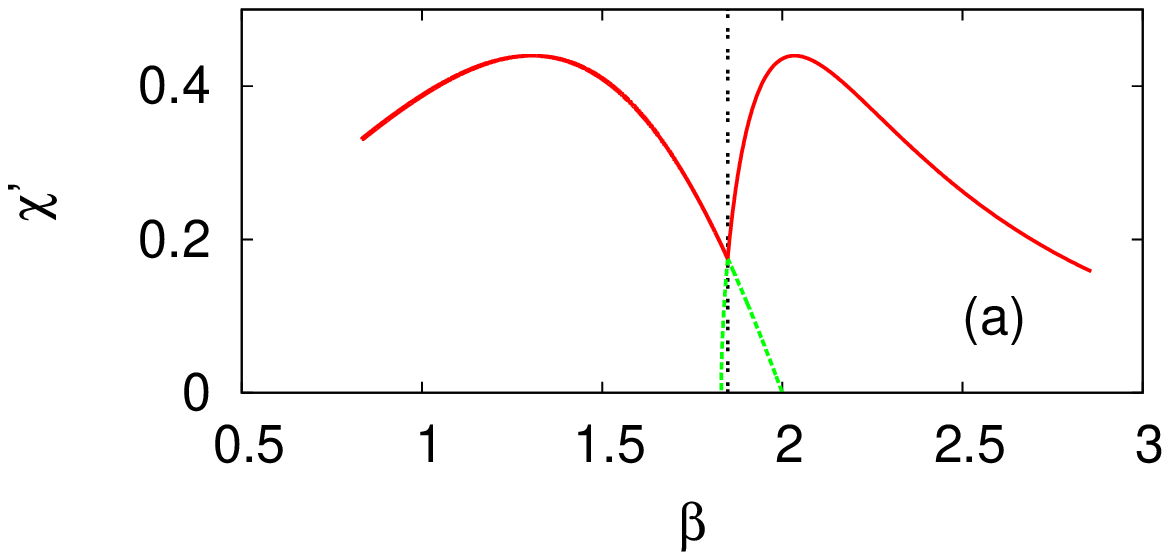}
\includegraphics[width=0.45\textwidth]{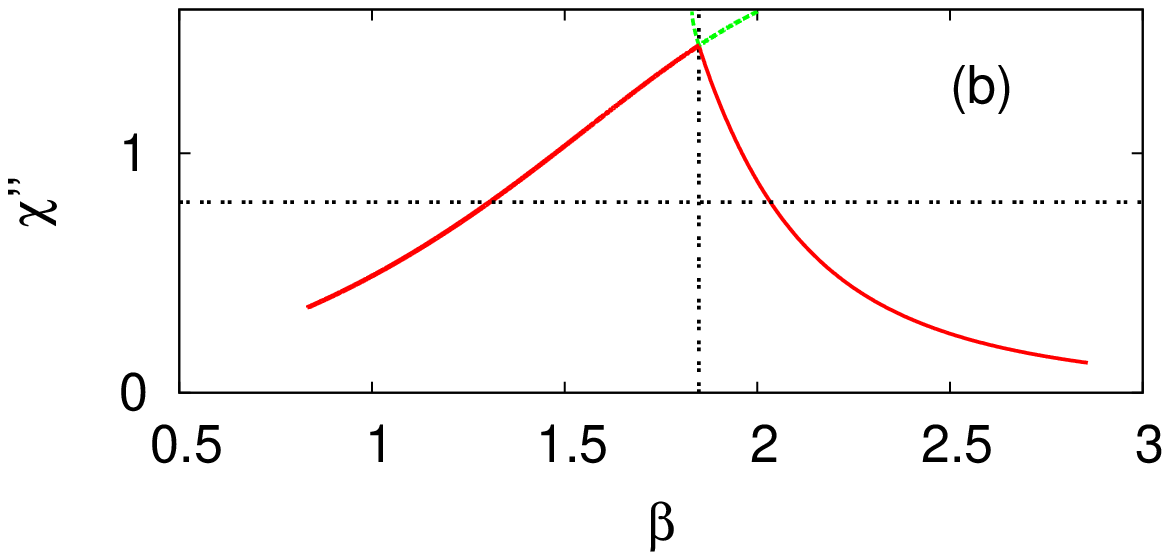}
\includegraphics[width=0.45\textwidth]{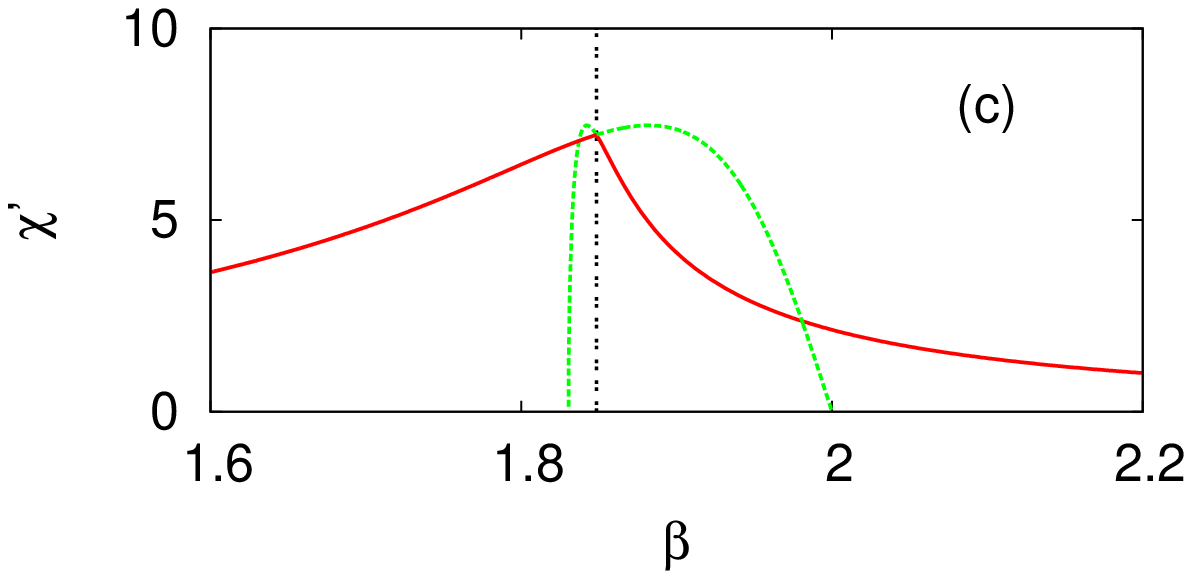}
\includegraphics[width=0.45\textwidth]{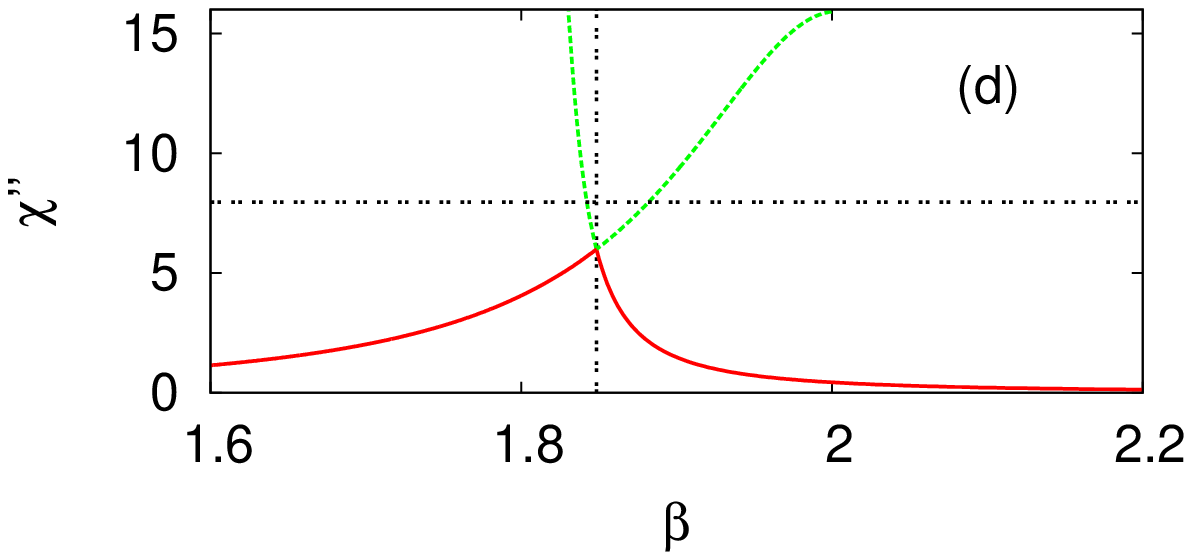}
\caption{Three-state clock model under the field in the $x$ direction.
(a)-(b)
ac susceptibilities at $\omega = 2\pi \times 10^{-1}$
and (c)-(d) the same quantities but at $\omega = 2\pi \times 10^{-2}$.
The solid lines represent values at the lowest free energy,
while the dotted lines mean what one can observe at metastable states. The
vertical dashed lines indicate the transition point $\beta_c =
\frac{8}{3} \ln 2 \approx 1.848~39$.
The horizontal dashed lines indicate $(2\omega)^{-1}$.
}
\label{fig:q3}
\end{figure}

\begin{figure}
\includegraphics[width=0.45\textwidth]{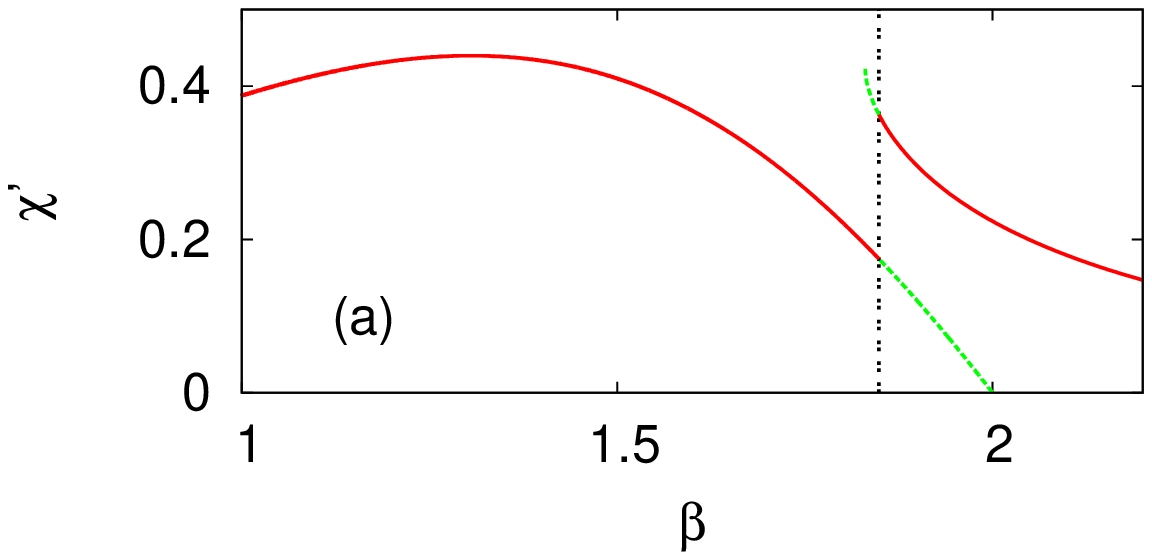}
\includegraphics[width=0.45\textwidth]{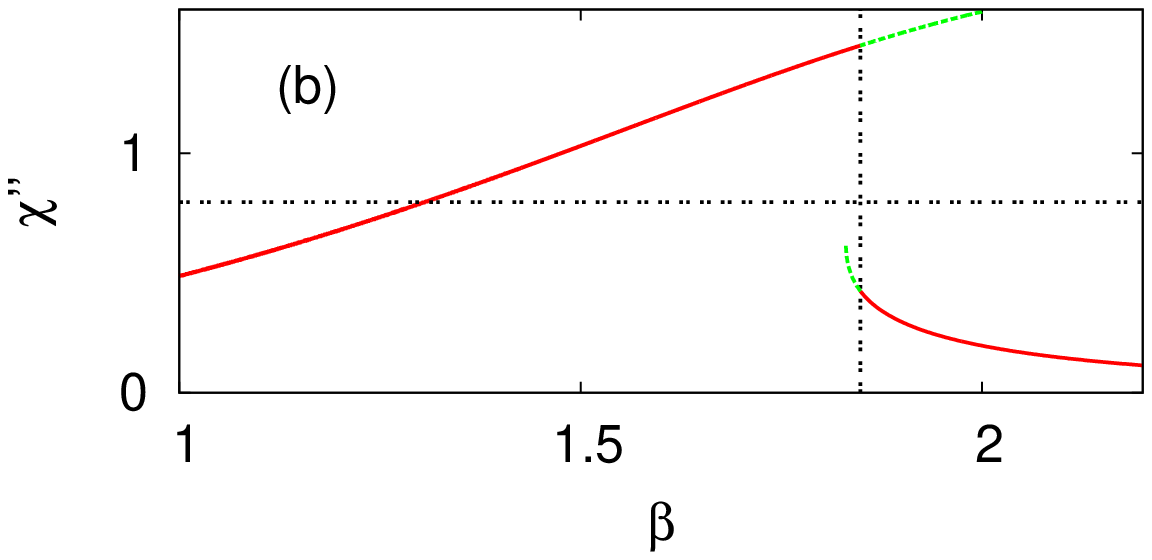}
\includegraphics[width=0.45\textwidth]{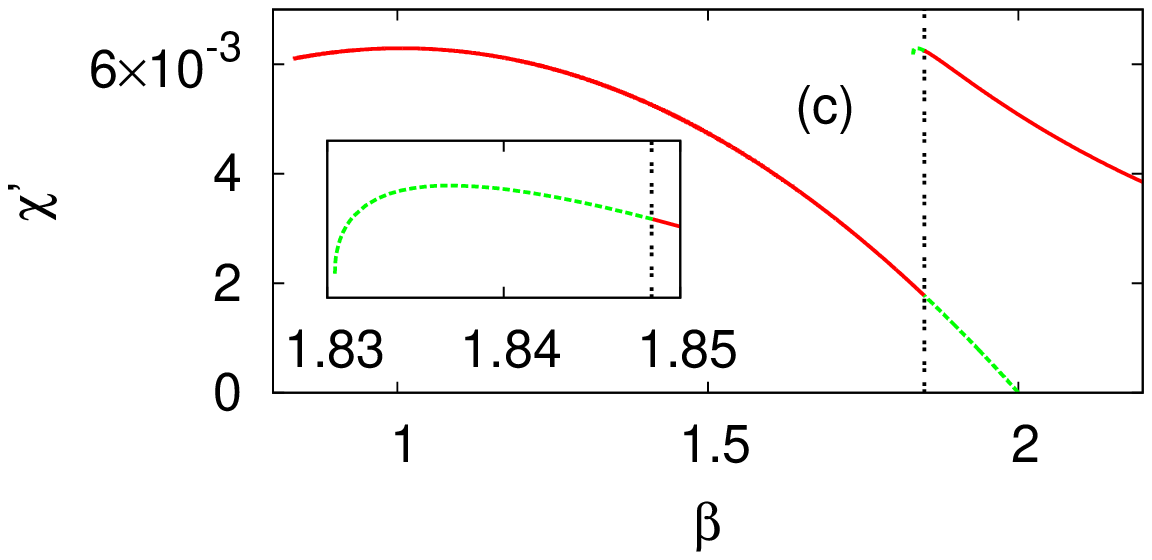}
\includegraphics[width=0.45\textwidth]{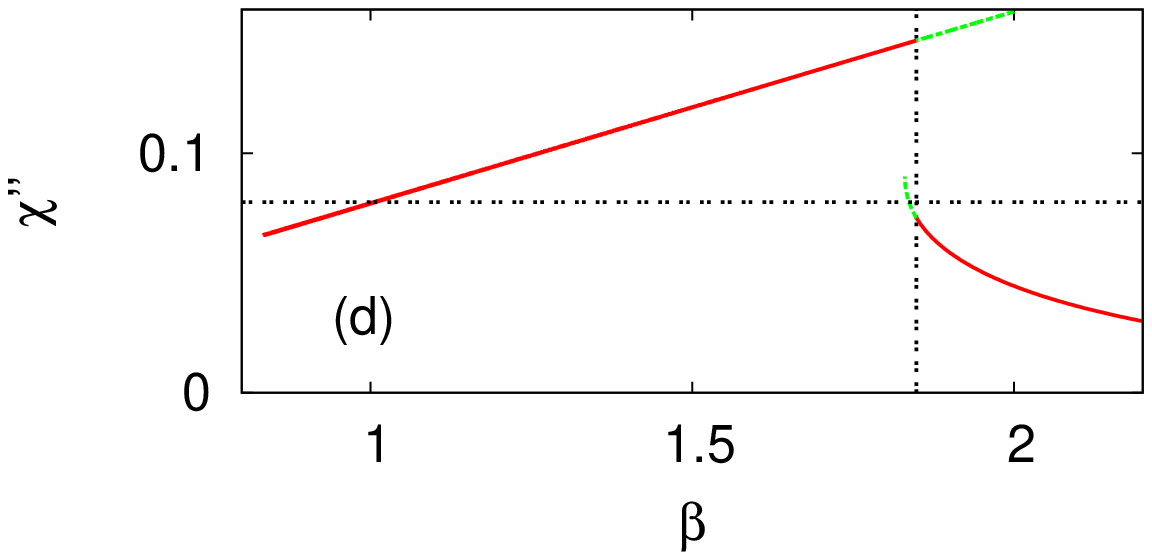}
\caption{Three-state clock model under the field in the $y$ direction.
(a)-(b)
ac susceptibilities at $\omega = 2\pi \times 10^{-1}$
and (c)-(d) the same quantities but at $\omega = 2\pi$. Inset: a zoomed view
of the left branch, showing a maximum around $\beta \approx 1.837$.
The solid lines represent values at the lowest free energy,
while the dotted lines mean what one can observe at metastable states. The
vertical dashed lines indicate the transition point $\beta_c =
\frac{8}{3} \ln 2 \approx 1.848~39$.
The horizontal dashed lines indicate $(2\omega)^{-1}$.
}
\label{fig:qy3}
\end{figure}

When the field is parallel to the magnetization,
one can infer from Fig.~\ref{fig:itauy3}(a) the possibility of
two resonance peaks only above a certain frequency since one can find two
temperatures where the external frequency matches the relaxation time scale
only if $\tau^{-1} > 1-\beta_c/2 \approx 0.075~803~8$. One cannot find proper
relaxation time scales to match the external driving frequency if
the frequency is too low. It is readily confirmed in
Figs.~\ref{fig:q3}(a) to \ref{fig:q3}(d). More precisely, the threshold of
$\omega$ for the stable double resonance peaks can be found by requiring
$\chi'' = (2 \omega)^{-1}$ to be met exactly at $\beta = \beta_c =
\frac{8}{3} \ln 2$. Since $\tau_{\parallel}^{-1}(\beta_c) = 1 - \beta_c / 2$, we get
the following quadratic equation,
\begin{equation}
\frac{\omega \tau_{\parallel} (\tau_{\parallel} - 1)}{1 + \omega^2 \tau_{\parallel}^2} =
\frac{1}{2\omega},
\label{eq:threshold}
\end{equation}
which yields $\omega = (3 - 4\ln 2)/\sqrt{24 \ln 2 - 9} \approx
0.0822986$.

When the field is perpendicular to the magnetization, one can guess that
the external frequency $\omega$ should be again large enough to find two
matching temperatures. In Fig.~\ref{fig:itauy3}(b), for example, the minimum
$\tau_{\perp}^{-1}$ above which there stably exist two matching temperatures is found to
be $\tau_{\perp}^{-1} = 1 - \frac{2}{3} \ln 2 \approx 0.537~902$.
It is true that one finds only one maximum in $\chi'$ when $\omega$ is low
[Figs.~\ref{fig:qy3}(a) and (b)].
However, our calculation shows that the double resonance is anyway impossible
if we take only truly stable states into account. The left maximum in
$\chi'$ can be located only on a metastable branch even for a very large
value of $\omega$ [Figs.~\ref{fig:qy3}(c)]. If we repeat the same
calculation as Eq.~(\ref{eq:threshold})
to find the threshold of $\omega$ with $\tau_{\perp}^{-1} = 1 - \frac{2}{3}\ln
2$, we indeed find that the equation does not possess any real solution,
confirming this impossibility.

\section{Summary}
\label{sec:summary}

In summary, we have studied stochastic resonance with the mean-field kinetic
version of the $q$-state clock model. The response under a periodic external
field now depends on the direction of the field relative to the
magnetization vector. When they are parallel, the double stochastic
resonance is observed for every $q > 3$ qualitatively in the same way as in
the kinetic Ising case ($q=2$)~\cite{double}. When the field is perpendicular to the magnetization
vector, on the other hand, the resonance peak is suppressed to a lower
temperature and eventually vanishes as $q \rightarrow \infty$ since the
$U(1)$ symmetry sets in. For $q=3$, the discontinuous transition should be
also taken into account, and we have concluded that the double resonance
peaks are observable in a truly stable manner only
when the external driving frequency is high enough and the field direction
is parallel to that of the magnetization vector.

\acknowledgments
B.J.K. was supported by the National Research
Foundation of Korea (NRF) grant funded by the Korea government (MEST) (No.
2010-0008758).

%

\appendix
\section{Derivation of Eq.~(\ref{eq:evolution})}
\label{app:eqm}
We multiply each side of Eq.~(\ref{eq:qmaster}) by an arbitrary function of
spin $l$, denoted by $f(\theta_l)$, and
carry out summation over all the possible configurations:
\begin{equation}
\sum_{\bm \theta}\frac{d}{dt}P({\bm \theta}; t) f(\theta_l)
= -\sum_{\bm \theta} \sum_{j=1}^N P({\bm \theta}; t) f(\theta_l)
+ \sum_{\bm \theta} \sum_{j=1}^N \sum_{\theta_j'}
w_j(\theta_j' \rightarrow \theta_j) P({\bm \theta}'; t) f(\theta_l) .
\label{eq:bracket}
\end{equation}
The sum over $j$ is decomposed into two parts: the sum over terms with $j \neq l$
and the term for $j=l$. Let us consider the former first.
If we write $\theta_j' = \theta_j + \Delta$, we get
\begin{eqnarray*}
&&-\sum_{\bm \theta} \sum_{j\neq l}^N P({\bm \theta}; t) f(\theta_l)
+ \sum_{\bm \theta} \sum_{j\neq l}^N \sum_{\Delta} 
w_j(\theta_j) P(\theta_1, \cdots, \theta_j + \Delta, \cdots,
\theta_N; t) f(\theta_l) \\
&=&-\sum_{\bm \theta} \sum_{j\neq l}^N P({\bm \theta}; t) f(\theta_l)
+ \sum_{\bm \theta} \sum_{j\neq l}^N \sum_{\Delta}
w_j(\theta_j-\Delta) P({\bm \theta}; t) f(\theta_l) = 0, 
\end{eqnarray*}
where $\sum_{\theta_j'} w_j(\theta_j') = 1$ has again been used.
With $\langle f(\theta_l) \rangle \equiv \sum_{ \bm{\theta}} P({\bm \theta}; t) f(\theta_l)$
and $\theta_l' = \theta_l + \Delta$ again, 
Eq.~(\ref{eq:bracket}) is now reduced to 
\begin{eqnarray*}
\frac{d}{dt} \left< f(\theta_l) \right> 
&=&-\left< f(\theta_l) \right> 
+ \sum_{\bm \theta} \sum_{\Delta} w_l(\theta_l) P(\theta_1, \cdots, \theta_l+\Delta, \cdots, \theta_N; t) f(\theta_l)\nonumber\\
&=&-\left< f(\theta_l) \right> 
+ \sum_{\bm \theta} \sum_{\Delta}
w_l(\theta_l - \Delta) P({\bm \theta}; t) f(\theta_l - \Delta)\nonumber\\
&=& -\left< f(\theta_l) \right> + \left< \sum_{\Delta} w_l(\theta_l-\Delta)
f(\theta_l - \Delta) \right>\nonumber\\
&=& -\left< f(\theta_l) \right> + \left<  \frac{\sum_{\Delta}
\exp[\beta F_l \cos(\theta_l-\phi_l-\Delta)]
f(\theta_l - \Delta)}{\sum_{\theta_l} \exp [\beta F_l \cos(\theta_l -
\phi_l)]} \right>\nonumber\\
&=& -\left< f(\theta_l) \right> + 
\left<  \frac{\sum_{\theta_l}
\exp[\beta F_l \cos(\theta_l-\phi_l)]
f(\theta_l)}{\sum_{\theta_l} \exp [\beta F_l \cos(\theta_l -
\phi_l)]} \right>.
\end{eqnarray*}

\section{Derivation of Eqs.~(\ref{eq:deltax}) and (\ref{eq:deltay})}
\label{app:delta}

The perturbation in $M_x$ is expanded up to the first order as follows:
\begin{eqnarray*}
\frac{d\delta_x}{dt}
&=& - M_x^\ast - \delta_x + \frac{\sum_\theta \cos\theta~
e^{\beta M_x \cos\theta} e^{\beta M_y \sin\theta}}{\sum_\theta
e^{\beta M_x \cos\theta} e^{\beta M_y \sin\theta}}\\
&=& - M^\ast - \delta_x + \frac{\sum_\theta \cos\theta~
e^{\beta M^\ast \cos\theta}
e^{\beta \delta_x \cos\theta} e^{\beta \delta_y \sin\theta}}
{\sum_\theta
e^{\beta M^\ast \cos\theta}
e^{\beta \delta_x \cos\theta} e^{\beta \delta_y \sin\theta}}\\
&\approx& - M^\ast - \delta_x + \frac{\sum_\theta \cos\theta~
e^{\beta M^\ast \cos\theta}
(1 + \beta \delta_x \cos\theta + \beta \delta_y \sin\theta)}
{\sum_\theta
e^{\beta M^\ast \cos\theta}
(1 + \beta \delta_x \cos\theta + \beta \delta_y \sin\theta)}\\
&=& - M^\ast - \delta_x + \frac{\sum_\theta \cos\theta~
e^{\beta M^\ast \cos\theta}
+ \sum_\theta \cos\theta~
e^{\beta M^\ast \cos\theta}
(\beta \delta_x \cos\theta + \beta \delta_y \sin\theta) }
{\sum_\theta
e^{\beta M^\ast \cos\theta}
+ \sum_\theta
e^{\beta M^\ast \cos\theta}
(\beta \delta_x \cos\theta + \beta \delta_y \sin\theta) }\\
&\approx& - M^\ast - \delta_x
+ \left\{ \sum_\theta \cos\theta~ e^{\beta M^\ast \cos\theta}
+ \sum_\theta \cos\theta~ e^{\beta M^\ast \cos\theta}
(\beta \delta_x \cos\theta + \beta \delta_y \sin\theta) \right\}\\
&&\times \left\{ \sum_\theta e^{\beta M^\ast \cos\theta}
- \sum_\theta e^{\beta M^\ast \cos\theta}
(\beta \delta_x \cos\theta + \beta \delta_y \sin\theta)\right\}
\times \left\{ \sum_\theta e^{\beta M^\ast \cos\theta} \right\}^{-2}\\
&=& - \delta_x
- \left\{ \sum_\theta \cos\theta~ e^{\beta M^\ast \cos\theta}
\sum_\theta e^{\beta M^\ast \cos\theta}
(\beta \delta_x \cos\theta + \beta \delta_y \sin\theta) \right\}
\times \left\{ \sum_\theta e^{\beta M^\ast \cos\theta} \right\}^{-2}\\
&&+ \left\{ \sum_\theta e^{\beta M^\ast \cos\theta}
\sum_\theta \cos\theta~ e^{\beta M^\ast \cos\theta}
(\beta \delta_x \cos\theta + \beta \delta_y \sin\theta) \right\}
\times \left\{ \sum_\theta e^{\beta M^\ast \cos\theta} \right\}^{-2}\\
&=& -\delta_x - \beta {M^\ast}^2 \delta_x
+ \frac{ \sum_\theta \cos\theta \exp(\beta M^{\ast} \cos\theta)
(\beta \delta_x \cos\theta + \beta \delta_y \sin\theta)}
{\sum_\theta \exp(\beta M^{\ast} \cos\theta)}.
\end{eqnarray*}
Likewise, we obtain
\[
\frac{d\delta_y}{dt} = -\delta_y
+ \frac{ \sum_\theta \sin\theta \exp(\beta M^{\ast} \cos\theta)
(\beta \delta_x \cos\theta + \beta \delta_y \sin\theta)}
{\sum_\theta \exp(\beta M^{\ast} \cos\theta)}.
\]
One can furthermore show that
\[ \sum_\theta \cos\theta \sin\theta \exp(\beta M^\ast \cos\theta) = 0, \]
by which we can rewrite the above equations as
\begin{eqnarray*}
\frac{d\delta_x}{dt} &=& -\left( 1 + \beta {M^\ast}^2 - \beta C \right) 
\delta_x,\\
\frac{d\delta_y}{dt} &=& -\left( 1 - \beta + \beta C \right)\delta_y,
\end{eqnarray*}
where
\[
C \equiv \frac{\sum_\theta \cos^2\theta \exp(\beta M^\ast \cos\theta)}
{\sum_\theta \exp(\beta M^\ast \cos\theta)}.
\]

\end{document}